\documentclass[]{article}  
 \usepackage{authblk}
\usepackage{amsmath}
\usepackage{amsfonts}
\usepackage{amssymb}
\usepackage[export]{adjustbox}
\usepackage{graphicx}
\usepackage{subfig}
\usepackage{floatrow}
\usepackage{color}
\usepackage{fullpage}

\usepackage[colorlinks=true, allcolors=blue]{hyperref}

\def\eps{\varepsilon}

\def\O{\scriptscriptstyle 0}
\def\epso{\eps_{\scriptscriptstyle 0}}
 
\def\lambdao{\lambda_{\scriptscriptstyle 0}}
\def\muo{\mu_{\scriptscriptstyle 0}}
\def\etao{\eta_{\scriptscriptstyle 0}}

\def\ko{k_{\scriptscriptstyle 0}}
\def\co{c_{\scriptscriptstyle 0}}

\def\lambdaomin{\lambda_{\O_{min}}}
\def\lambdaomax{\lambda_{\O_{max}}}

\def\##1{{\bf #1}}
\def\=#1{\underline{\underline{#1}}}

\def\*#1{\breve{{\bf #1}}}
\def\+#1{\breve{\underline{\underline #1}}}

\def\ux{\hat{\underline x}}
\def\uy{\hat{\underline y}}
\def\uz{\hat{\underline z}}

\def\.{\mbox{ \tiny{$^\bullet$} }}

\def\Nbar{\bar{N}}

\def\rel{_{rel}}

\def\JscOpt{\ensuremath{J_{SC}^{Opt}}}

\def\Rsph{R_{sph}}

\begin{document}

\begin{center}
\textbf{Optimization approach for  {optical absorption in three-dimensional structures including solar cells}}\\

\textit{Benjamin J. Civiletti$^a$,
Tom H. Anderson$^a$,
Faiz Ahmad$^b$,
Peter B. Monk$^a$, and
Akhlesh Lakhtakia$^b$}\\
$^a${University of Delaware, Department of Mathematical Sciences,
501 Ewing Hall, Newark, DE 19716, USA}
$^b${Pennsylvania State University, Department of Engineering Science and Mechanics, NanoMM--Nanoengineered Metamaterials Group,   University Park, PA 16802, USA}

\end{center}

\begin{abstract}
The rigorous coupled-wave approach (RCWA) and the differential evolution algorithm (DEA) were coupled
in a practicable approach to  {maximize absorption in} optical structures with three-dimensional morphology.
As a model problem,
 optimal values of four geometric parameters and the bandgaps
of three $i$-layers were
found   for an amorphous-silicon, multi-terminal, thin-film tandem solar cell comprising
three $p$-$i$-$n$ junctions with a metallic hexagonally corrugated back-reflector.
When  the optical short-circuit current density was chosen as the figure of merit   to be maximized,
only the bandgap of the topmost $i$-layer   was significant and the remaining six parameters played  minor roles.  {While this configuration would  absorb light very well, it would have poor electrical performance. This is because the optimization problem allows for the thicknesses and bandgaps of the semiconductor layers to change. We therefore devised another figure of merit that takes into account bandgap changes by estimating the 
open-circuit voltage.} The resulting configuration was found to be optimal with respect to all seven variable parameters.  
The RCWA+DEA optimization approach is applicable to other types of
photovoltaic solar cells as well as optical absorbers,  with the choice of the figure of merit being  vital to a successful outcome.
\end{abstract}

\section{INTRODUCTION}
\label{sec:intro}  
Three items are needed to numerically optimize the design of   {an optical absorber} such as a thin-film solar cell. The first item is a fast solver that can predict the performance of the device in a variety of configurations.  The second item is an optimization code that can mitigate the effect of local minima without excessive computational time.  The third item is  a figure of merit
   that adequately captures the 
desired performance characteristics of the device so that a good design will emerge.   Some of these choices 
 {are} explored in this paper.

The  rigorous coupled-wave approach (RCWA) \cite{MGPG,chateau_hugonin_1994} can be used to model
the optical performance of a thin-film  {optical absorbers}, as has been shown for solar cells
with  metallic back-reflectors that are periodically corrugated along one direction
\cite{thinfilm1,thinfilm2,solano_faryad_hall_mallouk_monk_lakhtakia_2013}.
Indeed, the RCWA provides accurate results with high computational speed
for boundary-value problems involving structures that are invariant only along, say,
the $y$ axis and therefore are quasi-two-dimensional \cite{gibbs,shuba1}. Furthermore,
the RCWA can be coupled with
 the differential evolution algorithm (DEA) \cite{storn_price_1997,price_storn_lampinen_2014} for optimization \cite{solano_faryad_hall_mallouk_monk_lakhtakia_2013}.
 However, the
computational requirements of RCWA increase significantly  when the back-reflector is periodically corrugated in two directions, i.e.,
the boundary-value problem is fully three-dimensional (3D) in nature \cite{Onishi,faiz}.
Additional design  {parameters  enter} the optimization process thereby to increase
the computational burden further.

The dimensions of the unit cell of an optical absorber with a PCBR directly affect optical absorption
\cite{solano_faryad_hall_mallouk_monk_lakhtakia_2013}.
Optimization  of thin-film solar cells with two-dimensionally corrugated back-reflectors 
 {for maximum absorption} has not been reported heretofore, to our knowledge. As a preliminary study showed
that it is becoming a practicable proposition with commonly available computational resources, we decided to implement the RCWA+DEA approach
to optimize a fully 3D   {absorbing} structure \cite{ben}.  

To demonstrate this approach, we report here the  {maximization of optical absorption in an idealized}
thin-film tandem solar cell fabricated over a periodically corrugated  
back-reflector (PCBR) with  hillock-shaped corrugations arranged on a hexagonal lattice. The active region of the chosen
solar cell comprises three electrically isolated  $p$-$i$-$n$ junctions.
 {The semiconductor layers were taken to have the bandgap-dependent optical properties of amorphous silicon.\cite{Ferlauto}}  Silver \cite{palik_1985}
is a good choice for the PCBR because  its plasmonic nature   can be harnessed to launch 
surface-plasmon-polariton (SPP) waves inside the device and thereby enhance the optical electric field and optical absorption \cite{Anderson,green_pillai_2012,FL2013}. With the foregoing choices,
our results
indicate that  {maximization of the
optical short-circuit current density, the standard figure of merit},\cite{Optim1,Optim2,Optim3} does not result in a desirable design. Instead, we found that
the maximum power density is a better figure of merit.
 
The plan of this paper is as follows. The optical boundary-value problem that is solved to determine  {the spectrally integrated number of absorbed photons per unit volume per unit time $N_{ph}$} is presented in Sec.~\ref{sec:boundary}. The numerical techniques adopted for this work are presented in Sec.~\ref{Sec:cm}:
the three-dimensional implementation of the RCWA is briefly described in Sec.~\ref{sec:rcwa}, Sec.~\ref{sec:eigen} contains the diagonalization of a matrix that emerges in the RCWA implementation, and Sec.~\ref{Sec:dea} briefly describes the DEA. Numerical results are provided in Sec.~\ref{sec:numerical}, Sec.~\ref{sec:converge} discusses the convergence of the numerical methods, while Sec.~\ref{sec:compare} briefly compares these numerical results to a thin-film tandem solar cell with a bi-sinusoidal PCBR. Closing remarks are presented in
Sec.~\ref{sec:close}.

The free-space wavenumber, angular frequency, and intrinsic impedance of free space are denoted by $k_{0}=2\pi/\lambdao$, $\omega=\ko\co$, and $\etao=\sqrt{\muo/\epso}$, respectively,  where  $\lambdao$ is the free-space wave length, 
$\muo$   is  the permeability of free space, 
 $\epso$ is  the   permittivity of free space,
and $\co=1/\sqrt{\epso\muo}$ is the speed of light in free space.  Vectors are underlined, column vectors and matrices associated with the RCWA are in boldface with breve notation, and the Cartesian unit vectors are identified as $\ux$, $\uy$, and $\uz$. The imaginary unit is denoted by $i=\sqrt{-1}$.

\section{MODEL BOUNDARY-VALUE  PROBLEM}
\label{sec:boundary}
\begin{figure} [ht]

   \begin{tabular}{cc} 
  \sidesubfloat[]{\includegraphics[height=8cm,width=8cm]{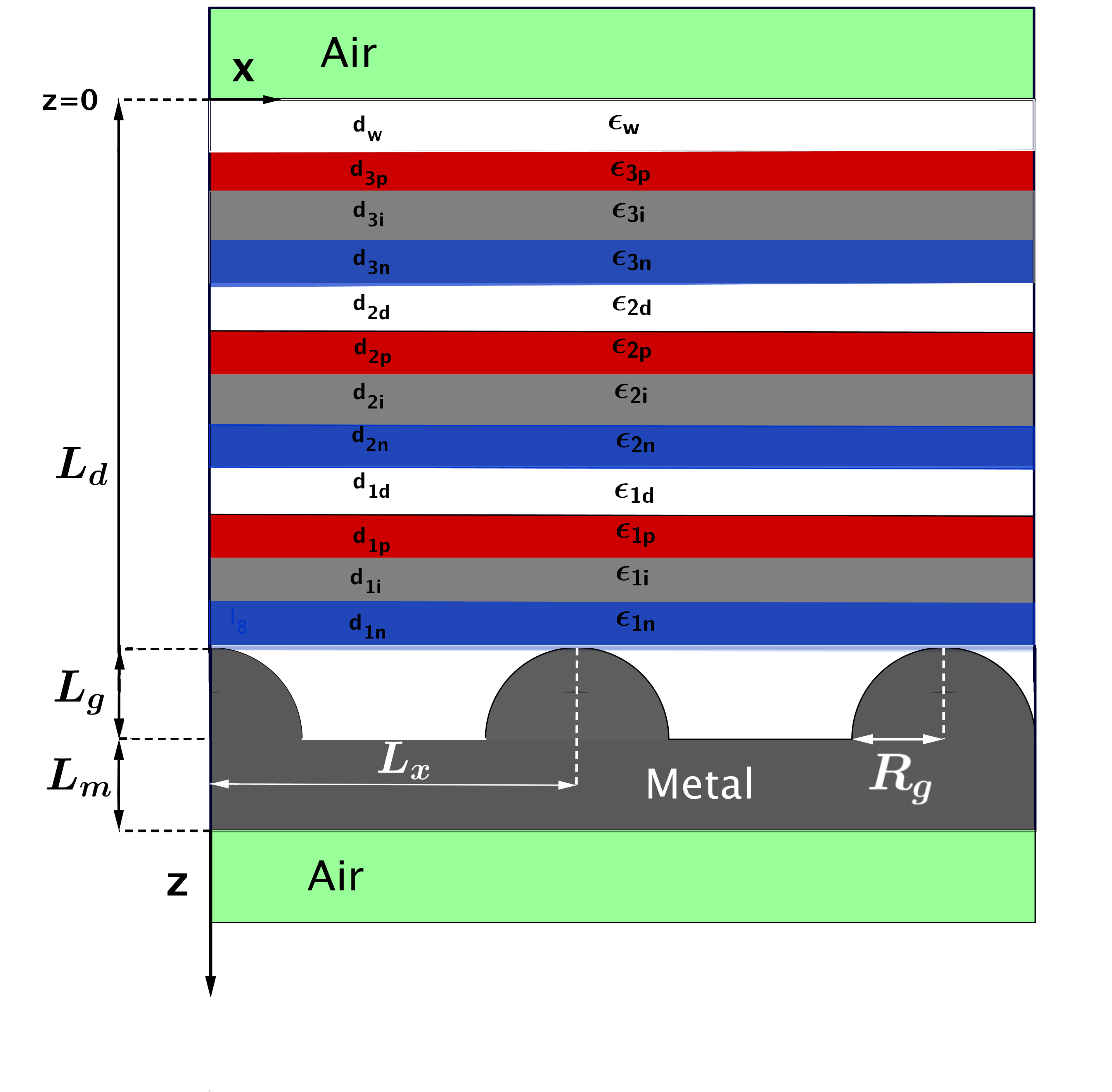}}
\sidesubfloat[]{
\includegraphics[height=7cm,width=5.06cm]{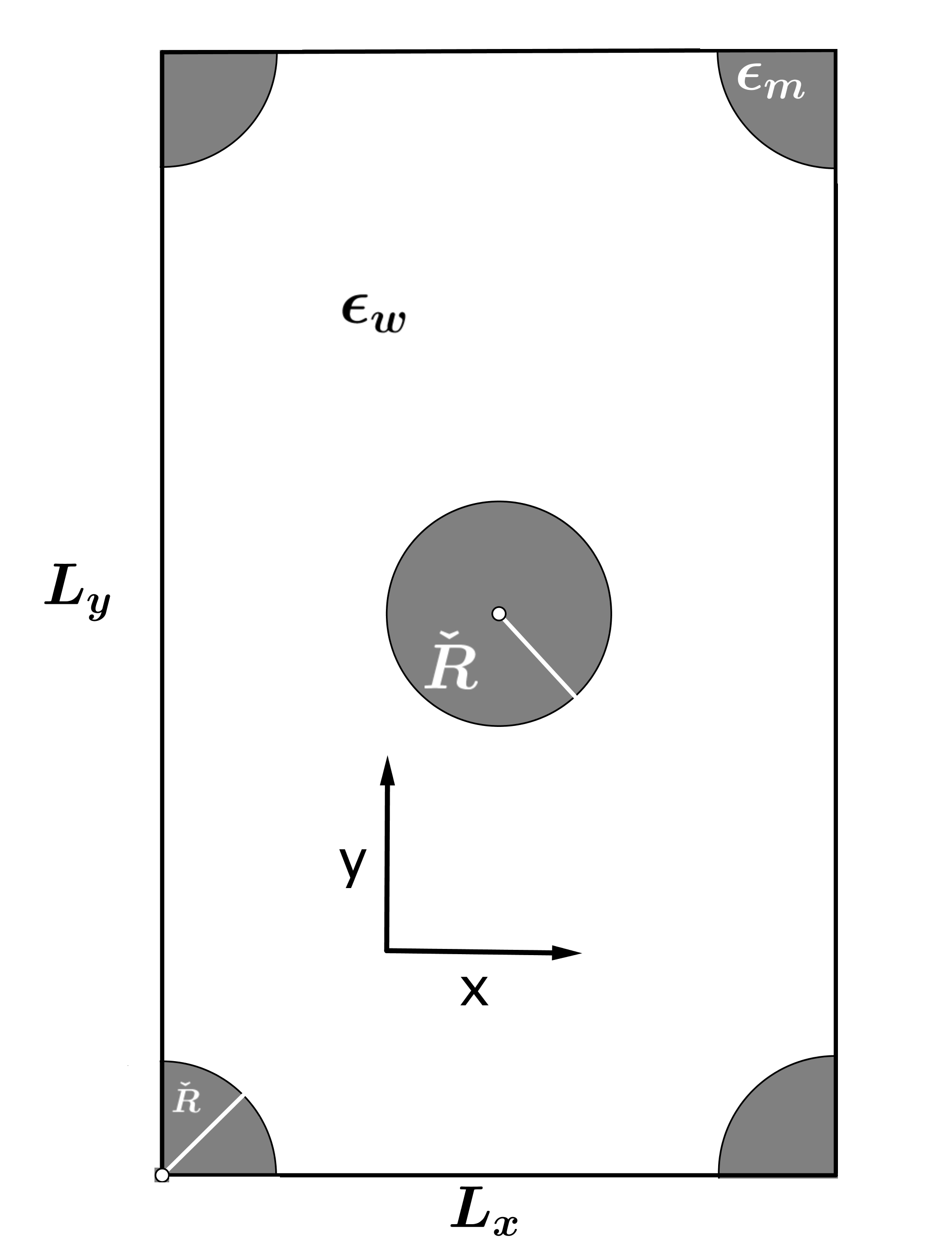}}
   
   \end{tabular}
   \caption{\label{fig:unit cell} (a) Schematic of the model boundary-value
   problem in the plane $y=0$.  The $n$-type semiconductor layers are blue, the $p$-type semiconductor layers are red, and the $i$-type are gray. White regions are occupied by a material with real relative permittivity $\eps_w(\lambdao)$. (b) Schematic of the grating region in the plane $z=z_g\in[L_d,L_d+L_g]$.
   } 
\end{figure}

We considered the  {boundary-value} problem shown schematically in Fig.~\ref{fig:unit cell}, which also defines the thicknesses $L_{d}, L_{g}$, and $L_{m}$. The  {device} occupies the region 
\begin{equation}
{\cal X}:\left\{(x,y,z)
\vert -\infty<x<\infty, -\infty<y<\infty, 0<z<L_d+L_g+L_m\right\},
\end{equation} with the half spaces $z<0$ and $z>L_d+L_g+L_m$ occupied by air.   The reference unit cell
is identified as  ${\cal R}:\left\{(x,y,z)
\vert -L_x/2<x<L_x/2,\right.$  $\left.-L_y/2<y<L_y/2,  0<z<L_d+L_g+L_m\right\}$, the back-reflector (which also functions as an electrode in a solar cell) being doubly periodic
with period $L_x$ along the $x$ axis and period $L_y$ along the $y$ axis.

The region $0< z < L_d$ comprises an antireflection window and three $p$-$i$-$n$ junctions which are electrically isolated from each other by two windows, as shown in Fig.~\ref{fig:unit cell}(a). The relative permittivity $\eps_d(x,y,z,\lambdao)$ of this multilayered material depends on  $\lambdao$.
The layers are identified in the figure. All windows are made of a material of  {relative}
permittivity $\eps_w(\lambdao)$. The $\lambdao$-dependent  {relative} permittivity of each semiconductor layer depends on the bandgap chosen for that layer.

The region $L_d +L_g < z < L_d+L_g+L_m$ is occupied by a metal with relative permittivity $\eps_m(\lambdao)$. 
The region $L_d < z < L_d+L_g$,
henceforth termed the grating region, contains a periodically undulating surface with period $L_x$ along the $x$ axis and   period $L_y$ along the $y$ axis.  The unit cell in the $xy$ plane was chosen to form a two-dimensional rectangular  lattice that is equivalent to a hexagonal lattice. If the side of the regular hexagons in this lattice is denoted by $L_h$, then  $L_x=L_h$ and $L_{y}=\sqrt{3}L_h$ for the rectangular lattice.

\floatsetup[figure]{style=plain,subcapbesideposition=top,font=bf}

The grating region is defined by  hillocks arranged as in 
Fig.~\ref{fig:unit cell}(b). Each hillock is a frustum of a sphere of radius $\Rsph$. The base of the hillock is a circle of radius ${R_g}$ and the height of the
hillock equals $L_g$, as shown in Fig.~\ref{fig:unit cell}(a); accordingly,
   \begin{equation}
\label{eq:sphere}
\displaystyle{
\Rsph=\frac{1}{2}\bigg(\frac{R_g^2}{L_g}+L_g\bigg)
} \, .
\end{equation}
The intersection of the plane $z=z_g\in[L_d,L_d+L_g]$ and a hillock is a circle of radius
\begin{equation}
\check{R}=\sqrt{(z_g-L_d)[2\Rsph-(z_g-L_d)]}\,.
\end{equation}
The relative permittivity $\eps_{g}(x,y,z,\lambdao)=\eps_{g}(x\pm L_x,y\pm L_y,z,\lambdao)$ in the grating region  is 
\begin{equation}
   \label{eq:perm}
\eps_{g}(x,y,z,\lambdao)=
      \eps_{m}(\lambdao)-[\eps_{m}(\lambdao)-\eps_{w}(\lambdao)] \ \mathcal{U}(x,y,z) \\
\end{equation}
with
\begin{equation}
   \label{eq:cap1}
\mathcal{U}(x,y,z)= \left\{
\begin{array}{ll}
      1, &  D\geq \check{R}\\
0, & D <\check{R}\\
\end{array} 
\right.\, 
\end{equation}
where  $D$ is the minimum distance between the point $(x,y,z)$
 and the centers $(0,0,z_{g})$, $(L_x,0,z_{g})$, $({L_x}/{2},{L_y}/{2},z_{g})$,
 $(L_x,L_y,z_{g})$ and $(0,L_y,z_{g})$ of the circle and four quarter circles in Fig.~\ref{fig:unit cell}(b).

\section{Numerical Techniques Used}\label{Sec:cm}
\subsection{Rigorous coupled-wave approach}
\label{sec:rcwa}
The RCWA was used the calculate the electric field phasor everywhere inside the  {chosen device} as a result of illumination by a monochromatic plane wave normally incident on the plane $z=0$ from the half space $z<0$. The electric field phasor of the incident plane wave was taken as
\begin{equation}
{\underline E}_{\rm inc}(x,y,z,\lambdao)=
{E_o} \frac{\ux+\uy}{\sqrt{2}}\exp\left(i\ko{z}\right)\,,
\end{equation}
where $E_{o}=4\sqrt{15 \pi}$~V~m$^{-1}$.

As a result of the PCBR being doubly periodic,
the $x$- and $y$-dependences of the electric and magnetic field phasors   {have
to be} represented everywhere by Fourier series as 
\begin{align}
\label{Fourier-E}
{\underline E}(x,y,z,\lambdao) &= \sum_{m=-\infty}^{m=\infty} 
\sum_{n=-\infty}^{n=\infty}
{\underline e}^{(m,n)}(z,\lambdao)\exp\left[i\left(k_x^{(m)}x+k_y^{(n)}y\right)\right]
\intertext{and}
{\underline H}(x,y,z,\lambdao) &= \sum_{m=-\infty}^{m=\infty} 
\sum_{n=-\infty}^{n=\infty}
{\underline h}^{(m,n)}(z,\lambdao)
\exp\left[i\left(k_x^{(m)}x+k_y^{(n)}y\right)\right],
\label{Fourier-H}
\end{align}
where $k_{x}^{(m)}= m({2\pi}/{L_x})$, $k_{y}^{(n)}= n({2\pi}/{L_y})$, and  ${\underline e}^{(m,n)} = {e}_x^{(m,n)}\ux+{e}_y^{(m,n)}\uy+{e}_z^{(m,n)}\uz$ as well as ${\underline h}^{(m,n)}= {h}_x^{(m,n)}\ux+{h}_y^{(m,n)}\uy+{h}_z^{(m,n)}\uz$ are  Fourier coefficients.
Likewise, the   {relative} permittivity $\eps\rel(x,y,z,\lambdao)$ everywhere is represented by the Fourier series
 \begin{equation}
\eps\rel(x,y,z,\lambdao) = \sum_{m=-\infty}^{m=\infty} 
\sum_{n=-\infty}^{n=\infty}
 {\eps}^{(m,n)}\rel(z,\lambdao)\exp\left[i\left(k_x^{(m)}x+k_y^{(n)}y\right)\right],
 \label{eq7}
\end{equation}
where  ${\eps}^{(m,n)}\rel(z,\lambdao)$ are Fourier coefficients.
Substitution of Eqs.~(\ref{Fourier-E})--(\ref{eq7}) in the frequency-domain Maxwell curl postulates yields the
matrix ordinary differential equation \cite[Chap.~3]{PMLbook}
\begin{equation}
\frac{d}{dz} \*f(z,\lambdao) =i\*P(z,\lambdao)\.\*f(z,\lambdao), 
\label{eq:ode}
\end{equation}  
where the column vector $\*f(z,\lambdao)$ contains the $x$- and $y$-directed components of the Fourier coefficients of the electric and magnetic field phasors.

Detailed descriptions of the  algorithm to solve Eq.~(\ref{eq:ode}) are available elsewhere \cite{ben,PMLbook}.
In brief, computational tractability requires the foregoing expansions to be truncated to include only  
$m \in \left\{-M_t,...,M_t\right\}$ and $n \in \left\{-N_t,...,N_t\right\}$, 
with $M_t\geq0$ and $N_t\geq0$.  The region $\cal R$ is partitioned into a sufficiently large number of slices along the $z$ axis. Each  slice is taken to be homogeneous along the $z$ axis but may be periodically nonhomogeneous along the $x$ and $y$ axes. Thus, the matrix $\*P(z,\lambdao)$ is assumed to be piecewise uniform in $z$. Boundary conditions are enforced on the planes $z=0$ and $z=L_d+L_g+L_m$ to match the fields to the incident, reflected, and transmitted  {fields}, as appropriate. A stable  marching algorithm is then used to determine the Fourier coefficients of the electric and magnetic field phasors in each slice\cite{PMLbook}.
Finally, the $z$-directed components of the electric and magnetic field phasors in the device can be obtained through algebraic equations arising during the derivation of Eq.~(\ref{eq:ode}).
 Thus, the electric field phasor ${\underline E}(x,y,z,\lambdao)$ can be determined throughout the solar cell.  

The spectrally integrated number of absorbed photons per unit volume per unit time is  given by
\begin{equation}
\label{eq20}
N_{ph}(x,y,z)= { \frac{1}{\hbar\co }}\int_{\lambdaomin}^{\lambdaomax}
\text{Im}\left\{\eps\rel(x,y,z,\lambdao) \right\} \left\vert
\frac{{\underline E}(x,y,z,\lambdao)}{E_o}
\right\vert^2
S(\lambdao) \,d\lambdao\,,
\end{equation}
where  
$\hbar$ is the reduced Planck constant,
$S(\lambdao)$ is the AM1.5G solar spectrum \cite{AM1.5},
${\lambdaomin} = 400$~nm,  and $\lambdaomax = 1100$~nm.  
With the assumption that the absorption of every photon in a semiconductor layer releases
an electron-hole pair, the
charge-carrier-generation rate   
 {$G(x,y,z)$ equals $N_{ph}(x,y,z)$,}
which can be determined at any location in the nine semiconductor layers.
The integral on the right side of
Eq.~(\ref{eq20}) was approximated using the trapezoidal rule
\cite{Jaluria} with the integrand uniformly sampled every $10$~nm.

\subsection{Diagonalization of $\*P(z,\lambdao)$}
\label{sec:eigen}
The numerically stable RCWA algorithm requires that $\*P(z,\lambdao)$ be diagonalizable \cite{Hochstadt} in each  slice for
every $\lambdao\in\left[\lambdaomin,\lambdaomax\right]$ \cite[Sec.~3.8]{PMLbook}. 
 As the relative permittivity is not uniform in any slice in the grating region, the 
built-in  function {\sf eig} in  Matlab\textsuperscript{\textregistered}
~(version R2016b) was used to compute the eigenvalues and
 eignenvectors of $\*P$. In all other slices, the eigenvalues and eigenvectors of $\*P$ were determined
 analytically, to increase the computational speed.

A superindex
\index{superindex}
\begin{equation}
\tau=m (2 N_t+1)+n \,,\quad
m\in\left[-M_t,M_t\right]\,,\quad
n\in\left[-N_t,N_t\right]\,,
\end{equation}
is defined for convenience, where $\tau_t=2M_tN_t+M_t+N_t$.\cite{PMLbook}
In any slice in which the  {relative} permittivity $\eps\rel(x,y,z,\lambdao)$ is  independent of
$x$ and $y$, the $2(2\tau_{t}+1)$ distinct
eigenvalues of $\*P$ are given by 
\begin{equation}
\label{eq:eigenvalues}
g_{\tau}^{\pm} = \pm \sqrt{ {\ko^{2}\eps\rel}
-\big(\breve{k}_{x}^{(\tau)}\big)^2-\big(\breve{k}_{y}^{(\tau)}\big)^2} \,,\quad
\tau\in\left\{-\tau_t,\cdots,\tau_{t}\right\}\,,
\end{equation}
with each eigenvalue having an geometric multiplicity of $2$,
$\breve{k}_x^{(\tau)} = k_x^{(m)}$, and
$\breve{k}_y^{(\tau)} = k_y^{(n)}$.
Half of the $4(2\tau_{t}+1)$ eigenvectors are 
\begin{eqnarray}
\nonumber
\mathbf{v}_{\tau}^{1\pm} &=& \left[0,\cdots,0,
\left({g_{\tau}^{\pm}}\right)^{-1}\left\{
\omega \muo-\frac{1}{\omega {\epso\eps\rel}}\left(\breve{k}_{x}^{(\tau)}\right)^2\right\}
,
0, \cdots,0,
-\frac{\breve{k}_{x}^{(\tau)} \breve{k}_{y}^{(\tau)}}{{\omega {\epso\eps\rel}}g_{\tau}^{\pm}}
,
0,\cdots,0,1,0,\cdots,0\right]^{T} \,,\\[5pt]
&&\qquad\qquad\qquad
\tau\in\left\{-\tau_t,\cdots,\tau_{t}\right\}\,.
\label{eq:eigenvalues1}
\end{eqnarray}
In the column vector on the right side of Eq.~(\ref{eq:eigenvalues1}),
 the non-zero entries occur  in the following locations (counting from the top):
$\tau$, $\tau+2\tau_{t}+1$, and $\tau+6\tau_{t}+3$.
The remaining $2(2\tau_{t}+1)$ eigenvectors are 
\begin{eqnarray}
\nonumber
\mathbf{v}_{\tau}^{2\pm} &=& \left[0,\cdots,0,
\frac{\breve{k}_{x}^{(\tau)} \breve{k}_{y}^{(\tau)}}{{\omega {\epso\eps\rel}}g_{\tau}^{\pm}}\ ,
0, \cdots,0,
-
\left({g_{\tau}^{\pm}}\right)^{-1}\left\{
\omega \muo-\frac{1}{\omega {\epso\eps\rel}}\left(\breve{k}_{y}^{(\tau)}\right)^2\right\}
,
0,\cdots,0,1,0,\cdots,0\right]^{T}\,\\[5pt]
&&\qquad\qquad\qquad
\tau\in\left\{-\tau_t,\cdots,\tau_{t}\right\}\,.
\label{eq:eigenvalues2}
\end{eqnarray}
The non-zero entries occur  in the column vector
 on the right side of Eq.~(\ref{eq:eigenvalues2}) in the following locations (counting from the top):
$\tau$, $\tau+2\tau_{t}+1$, and $\tau+4\tau_{t}+2$. 
 
\subsection{Differential evolution algorithm}\label{Sec:dea}
We employed the DEA to maximize a figure of merit over a variety of  optical and electrical parameters
numbering $\Nbar$. The DEA has been used previously\cite{solano_faryad_hall_mallouk_monk_lakhtakia_2013} for seeking optimal designs of PCBRs that are invariant along the $y$ axis, as the algorithm is well-suited to search a large space of candidate solutions. The number of candidate solutions depends on the number of parameters and discretization of parameter ranges for the optimization. 

To maximize the figure of merit $C:\mathcal{S} \subset \mathbb{R}^{\Nbar}\to \mathbb{R}$, we wish to find an optimal point $\mathcal{V}^{opt}\equiv\{v_{1},v_{2},\cdots,v_{\Nbar} \}\in \mathcal{S}$, where $\mathcal{S}$ is a search space of all possible parameter combinations.  
 
 We note that the DEA requires that all parameter ranges be discretized, so  the search is conducted over a finite number of possible outcomes. The DEA requires  specification of the crossover probability $C_R\in (0,1)$, a differential weight $\alpha \in (0,2)$ and  the number of random points $N_P$. Details of the algorithm have been provided elsewhere\cite{solano_faryad_hall_mallouk_monk_lakhtakia_2013,storn_price_1997,price_storn_lampinen_2014}.

The DEA is very useful for solving complicated optimization problems, but it does not guarantee convergence to a global extremum~\cite{deaconv}. However, stochastic sampling of the search space helps to avoid local maxima.

\section{NUMERICAL RESULTS AND DISCUSSION}
\label{sec:numerical}
In this paper  {we   have chosen to maximize optical absorption}.  The figures of merit defined later in this section take into account all optical effects such as  {the excitation of} SPP waves and waveguide modes \cite{Liu2016}.  This allows a tradeoff between the various optical phenomena without prejudicing one mechanism over another.  However, the parameter space is chosen so
that  {the excitation of SPP waves} and waveguide modes can be supported.

For all numerical results in this paper, the window layers were chosen to be made of alumimum-doped zinc oxide (AZO). For the two windows between
$p$-$i$-$n$ junctions, the thicknesses were fixed so that
$d_{2d}=d_{1d}=20$~nm. The  {relative} permittivity $\eps_w(\lambdao)$ of AZO was taken from a standard source\cite{xiaoyong_yan_qing-geng_2010}. The minimum thickness
of the PCBR was fixed at $L_m=150$~nm. The metal was chosen to be silver, whose  
 {relative} permittivity $\eps_m$ also depends on $\lambdao$\cite{palik_1985}. 

The  bandgaps ${\sf E}_{\ell{i}}$, $\ell\in\{1,2,3\}$, of the $i$-layers in the triple-junction solar cell were kept variable in the range $[1.3,1.95]$ eV, 
but their thicknesses $d_{1i}=d_{2i}=d_{3i}=200$~nm were kept fixed.
The thicknesses of all three $n$-layers and all three $p$-layers were also kept
fixed: $d_{\ell{n}}=d_{\ell{p}}=20$~nm. The bandgaps of all three $n$-layers
were fixed as ${\sf E}_{\ell{n}}=1.8$~eV, $\ell\in\{1,2,3\}$. The bandgaps
of the $p$-layers were fixed as follows: ${\sf E}_{1p}=1.8$~eV and
${\sf E}_{2p}={\sf E}_{3p}=1.95$~eV.  

By introducing C or Ge into the lattice, a new material is formed, but the $i$-layers are still of the a-Si:H\_GeC family. This process changes the bandgap, where the $\lambdao$-dependent  {relative  permittivity of the material is} obtained by an analytical model.\cite{FL2013,Ferlauto}. The electrical properties of the material also change, and can be found by applying Vegard's law to known values.\cite{Vegard} Since the electrical properties have no  {effect} on our optical model, they have no role in this study.

 Furthermore, the lattice parameter $L_h\in[200,800]$~nm, the antireflection window's thickness $d_w\in[10,130]$~nm, the base radius ${R_g}\in[10,400]$~nm, and the corrugation height $L_g\in[0,300]$~nm  were allowed to vary.
  Thus, the dimension of the search space $\cal S$ was $\Nbar=7$,
 and we sought an optimal design over a candidate space of $9 \times 10^{11}$ possible configurations. We used parameter values $C_R=0.7$, $\alpha=0.8$, and $N_P=70$ for
 optimization.

\subsection{Optimization for optical short-circuit current density}\label{Sec:4.1}
 {The figure of merit $C$ for the DEA optimization was initially chosen to be standard figure of merit for optical modeling of solar cells\cite{FOM}:
the optical short-circuit current density}
\begin{equation}
   \label{eq:total}
\JscOpt=\frac{q_e}{L_{x}L_{y}} \iiint_{{\cal R}_{sc}} {N_{ph}(x,y,z)}\,dx\,dy\,dz  \,,
\end{equation}
where ${{\cal R}_{sc}}$ is the portion of the reference unit cell $\cal R$
occupied by the nine semiconductor layers in the solar cell.
and $q_{e}=1.6 \times 10^{-19}$~C is the elementary charge.  {This figure of merit will maximize the number of photons absorbed in the solar cell, but disregards all electrical properties. In this section we proceed to show that this results in a poor design.}
\begin{figure}
\caption{\label{fig:DEA1} 
\JscOpt~in relation to (a) ${\sf E}_{3i}$, (b) ${\sf E}_{2i}$, and (c)  {${\sf E}_{1i}$.}
Each marker ($\mathbf{\cdot}$) represents a choice of parameters by DEA as the algorithm progresses. Larger values of $\JscOpt$ are desirable.
}
\begin{tabular}{cc}
  \includegraphics[width=70mm]{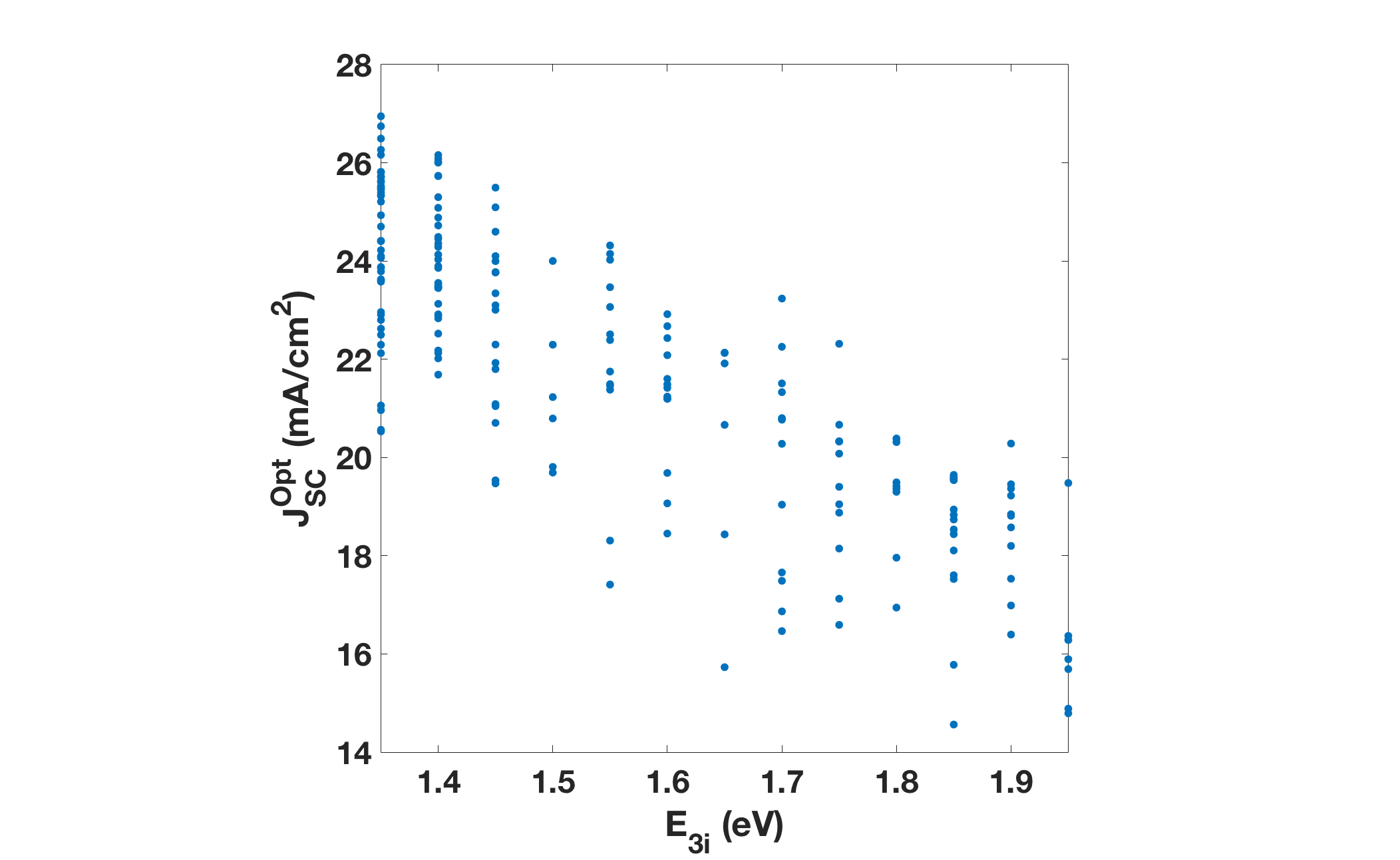} &   \includegraphics[width=70mm]{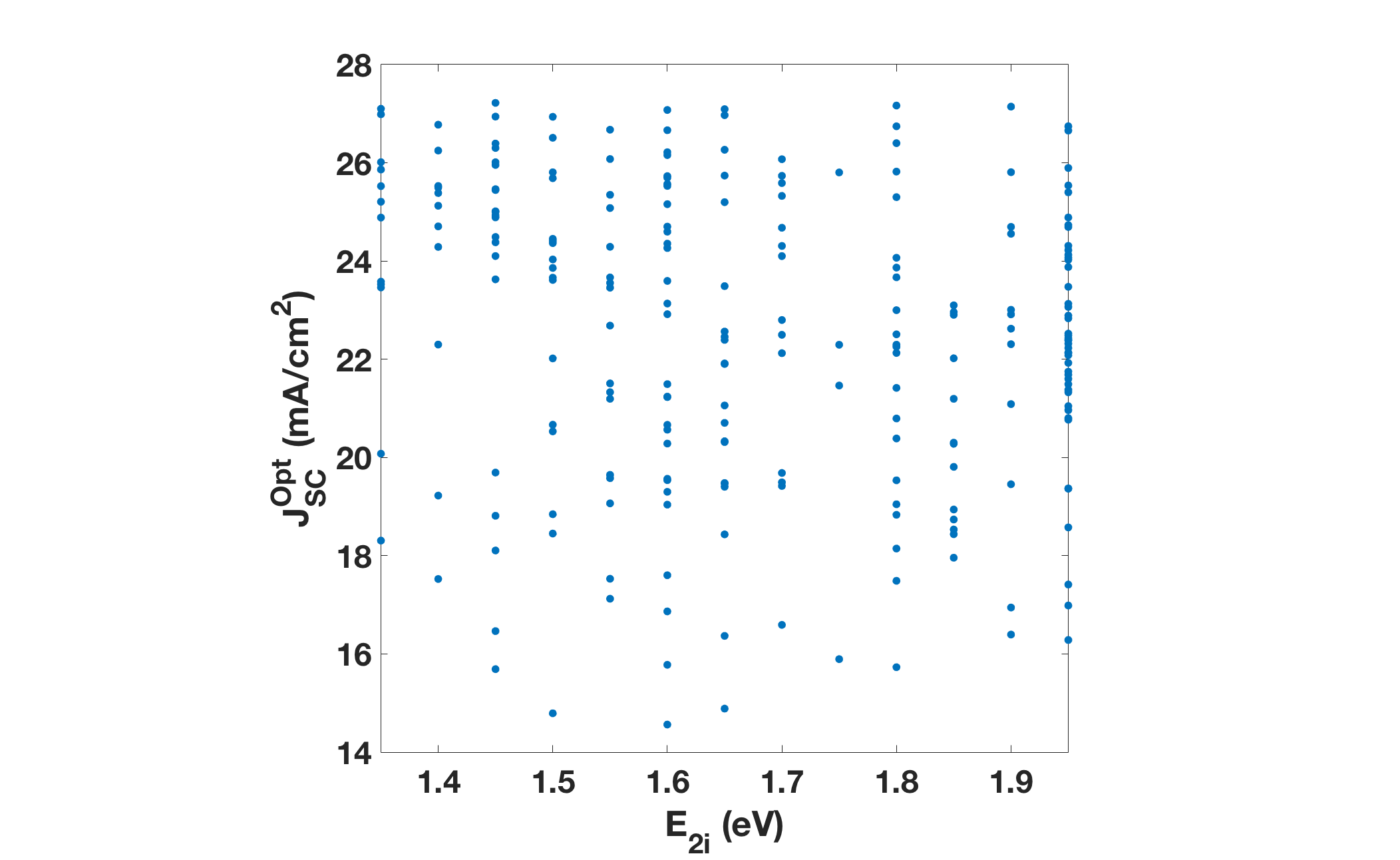} \\
(a)  & (b)  \\[6pt]
 \includegraphics[width=70mm]{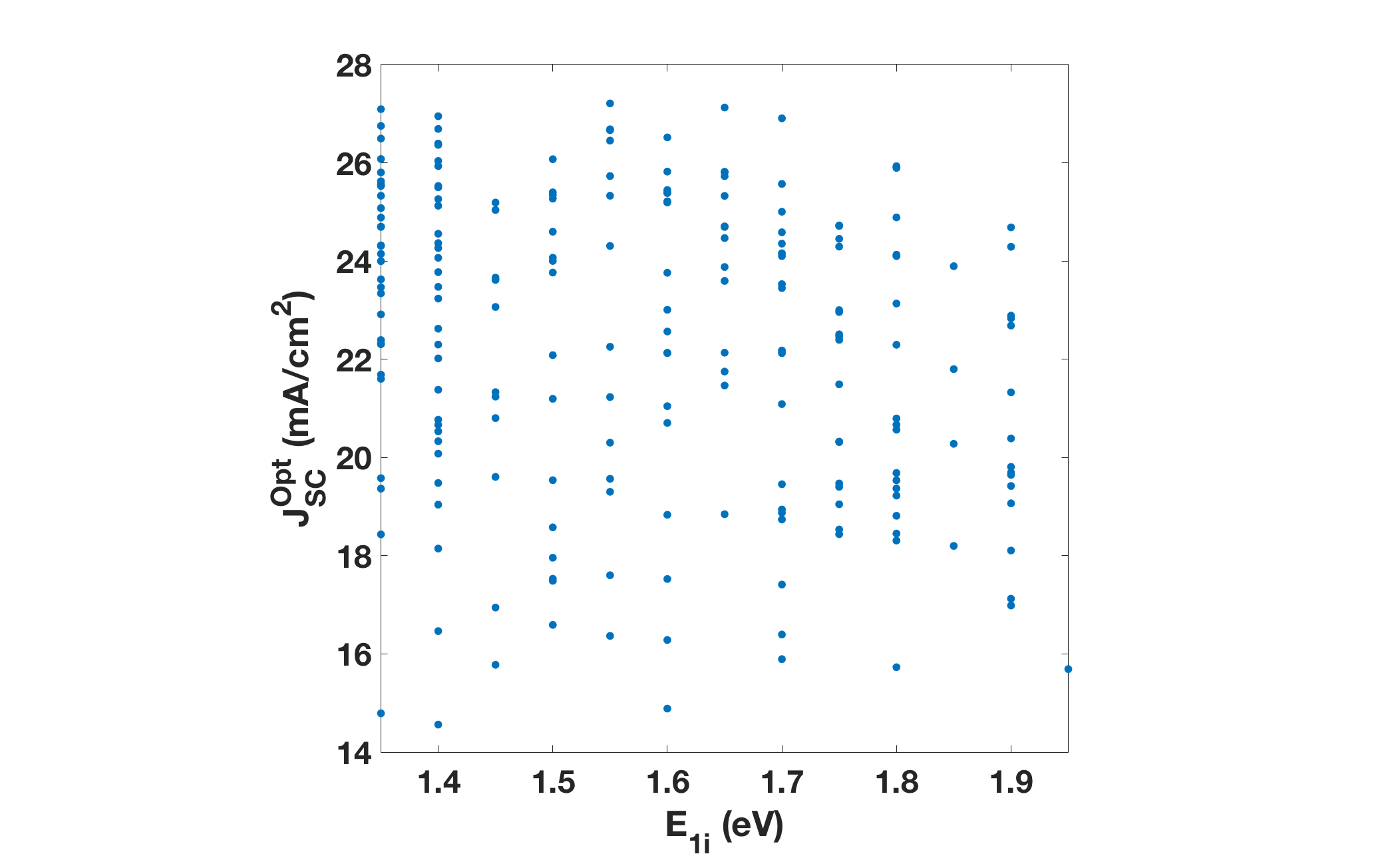} & \\
(c)  \\[6pt]
\end{tabular}
\end{figure}

Plots of $\JscOpt$ against the bandgaps ${\sf E}_{\ell{i}}$, $\ell\in\{1,2,3\}$,
for points sampled by the optimization exercise are presented in Fig.~\ref{fig:DEA1}. 
In this figure as well as in Fig.~\ref{fig:DEA2}, the data points from DEA are projected onto the plane formed by the variable being investigated and the figure of merit.
The bandgap of the topmost $i$--layer is the most influential parameter that controls $\JscOpt$. While attempting to maximize $J_{SC}^{Opt}$, the DEA minimized the bandgap in this layer, and we see that many parameter sets with ${\sf E}_{3i}^{opt}=1.3$~eV at the boundary of the constraint set were evaluated. As  ${\sf E}_{3i}$ increases from $1.3$~eV to $1.95$~eV, $\JscOpt$ decreases throughout most of this interval as shown in Fig.~\ref{fig:DEA1}(a). 

Maximization of $\JscOpt$ showed that minimizing ${\sf E}_{3i}$ was the most important factor compared with the remaining $\Nbar-1$ variable parameters (i.e., ${\sf E}_{1i}$, ${\sf E}_{2i}$, $L_h$, $d_w$,
${R_g}$, and $L_g$).
This phenomenon is clearly exemplified in Fig.~\ref{fig:DEA1}(a), where for any fixed value of ${\sf E}_{3i}$, the remaining $\Nbar-1$ variable parameters  contribute only to a $\pm 2$~mA~cm$^{-2}$ variation in  $\JscOpt$. Unfortunately, such a configuration is likely to be electrically inefficient: with a narrow bandgap, more charge carriers are excited, but the  operating voltage of the solar cell will be reduced.~\cite{physics}
Contrast this to  Figs.~\ref{fig:DEA1}(b) and (c) wherein the variations of $J_{SC}^{Opt}$ with the bandgaps ${\sf E}_{2i}$ and 
${\sf E}_{1i}$, respectively,   are  shown.  We see that while $J_{SC}^{Opt} $ is maximum when ${\sf E}_{1i}={\sf E}_{2i}=1.35$~eV, there are values of $J_{SC}^{Opt}$ ranging from $18 $~mA~cm$^{-2}$ to $27 $~mA~cm$^{-2}$.

Parenthetically, when the thicknesses of the $i$-layers were included as variables in an optimization exercise, the DEA simply focused on the maximization of those  thicknesses. 
The resulting configuration would also have poor electrical performance. While this type of solar cell would absorb more light, an excited charge carrier would have to travel further to reach an electrode, thereby increasing recombination and decreasing  efficiency.  We note that Fig.~\ref{fig:DEA1} demonstrates the choice of figure of merit is very important. We only  {included} these results to contrast them with numerical results in Fig.~\ref{fig:DEA4}, since all the chosen parameters in the optimization exercise should affect the solar-cell performance.

\subsection{Optimization for maximum power density} 
\label{sec:weight}
In order to improve the  optimal design  {of the chosen solar cell without including a full electrical model}, we 
devised a new figure of merit for the DEA which penalizes the effect of minimizing the bandgap of any of the $i$-layers. We defined the power density 
\begin{equation}
\displaystyle{
P_{sup}= \frac{1}{L_{x}L_{y}} \sum_{\ell=1}^3 {\sf E}_{\ell{i}}
 \iiint_{{\cal R}_{\ell}} {N_{ph}(x,y,z)}\,dx\,dy\,dz} 
 \label{eq:psup}
 \end{equation}
 as the new figure of merit, with ${{\cal R}_{\ell}}\subset{\cal R}$ being the region
 occupied by the $\ell$-th $p$-$i$-$n$ junction. Let us note
 that $P_{sup}$ is a theoretical upper bound on the maximum extractable power density of the solar cell.
 Furthermore, the summation over the index $\ell$ indicates that the tandem solar cell is to be
 configured in the multi-terminal format. Note that this power density is computed solely from the absorption of photons and the material bandgap. This estimates the maximum electrical power density but does not
 involve any electrical modeling  {(e.g., recombination and mobility of electrons and holes)}.   A useful extension of our approach would be to include an electrical model, but  {that extension lies beyond the scope of this} paper.

\begin{figure}
\caption{\label{fig:DEA2} 
$P_{sup}$
in relation to (a) $L_h$, (b) $d_w$, (c) $\zeta={R_g}/L_h$, and $L_g$.
Each marker ($\mathbf{\cdot}$) represents a choice of parameters by DEA as the algorithm progresses. The red marker indicates the maximum value of $P_{sup}$ achieved. }
\begin{tabular}{cc}
  \includegraphics[width=70mm]{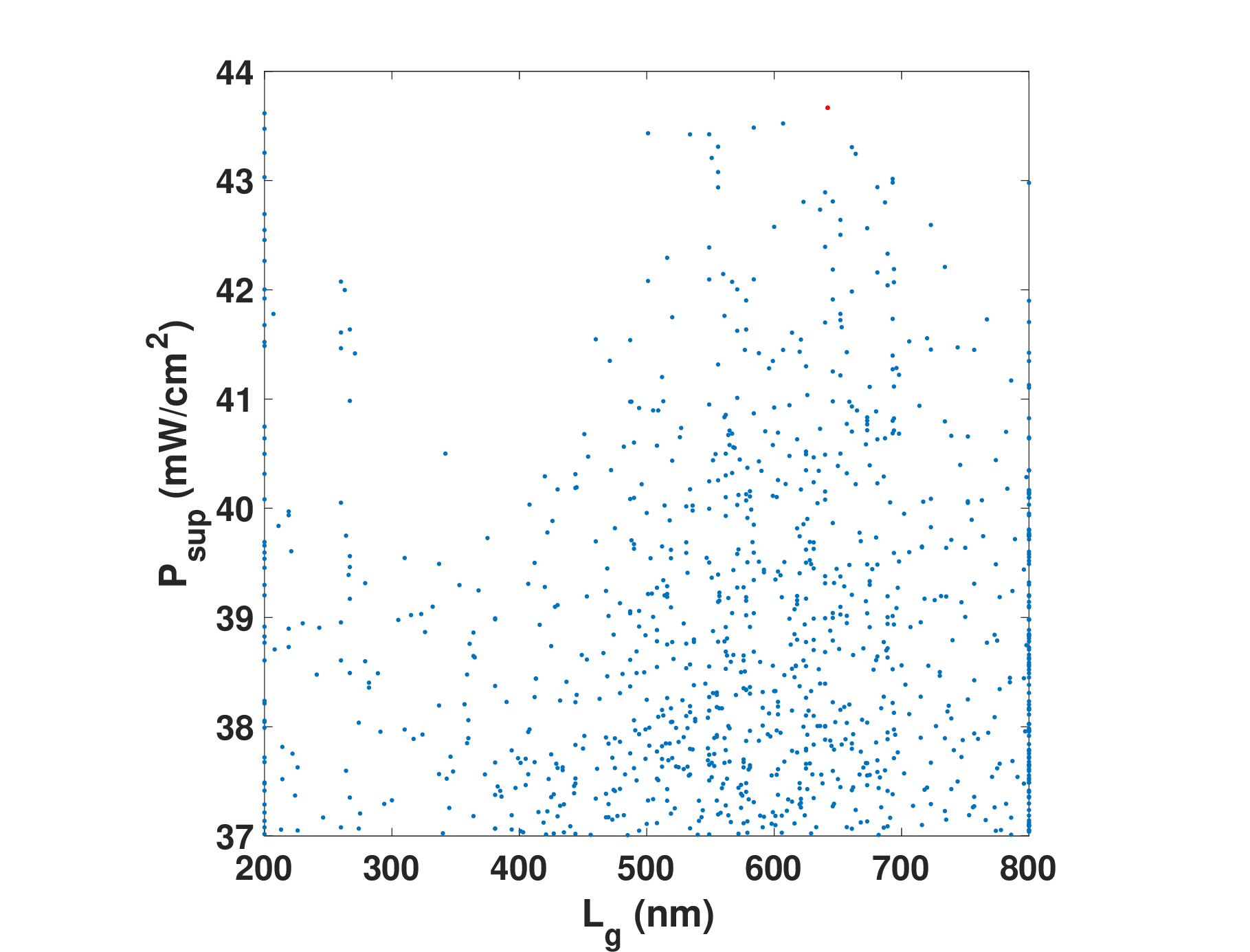} &   \includegraphics[width=70mm]{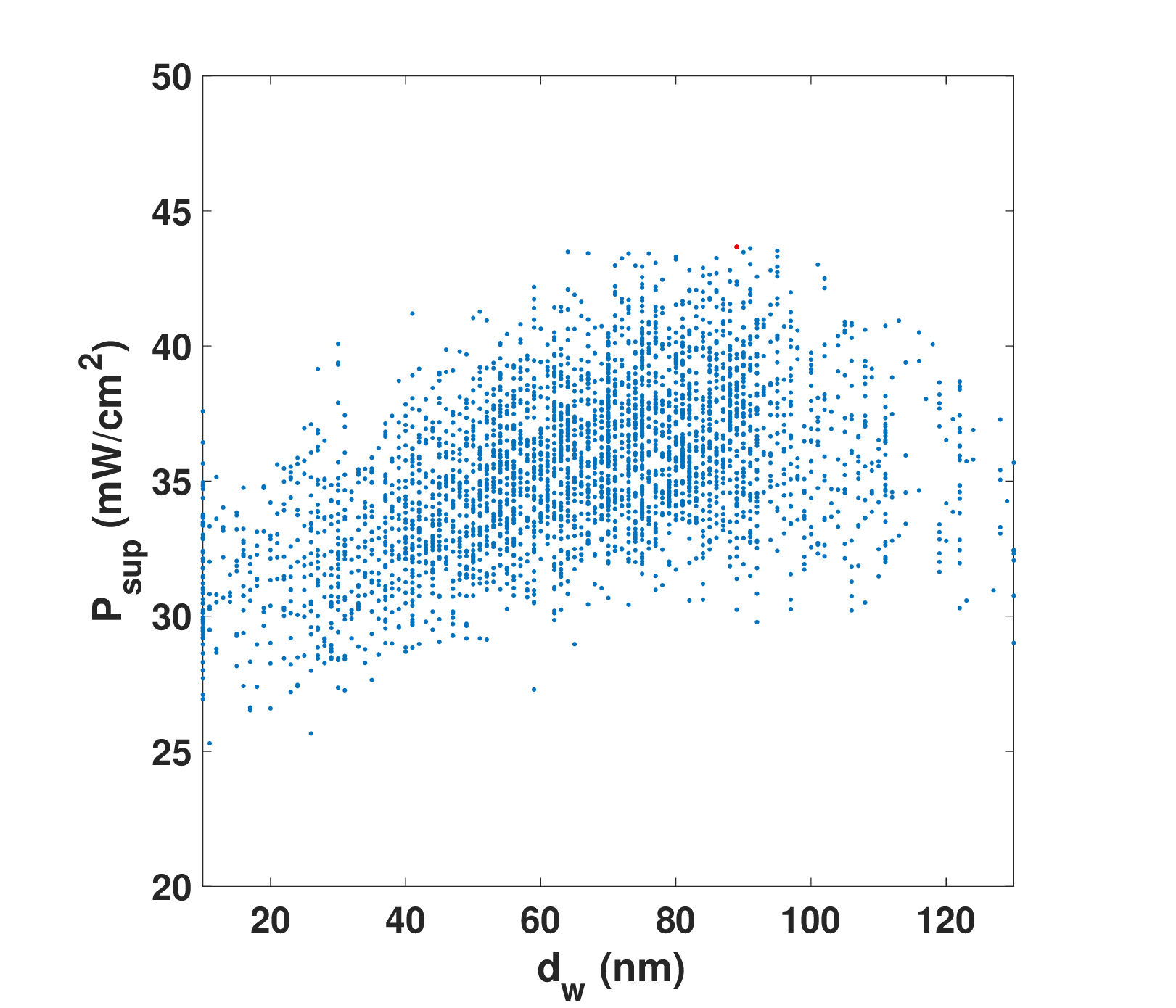} \\
(a)  & (b)  \\[6pt]
 \includegraphics[width=70mm]{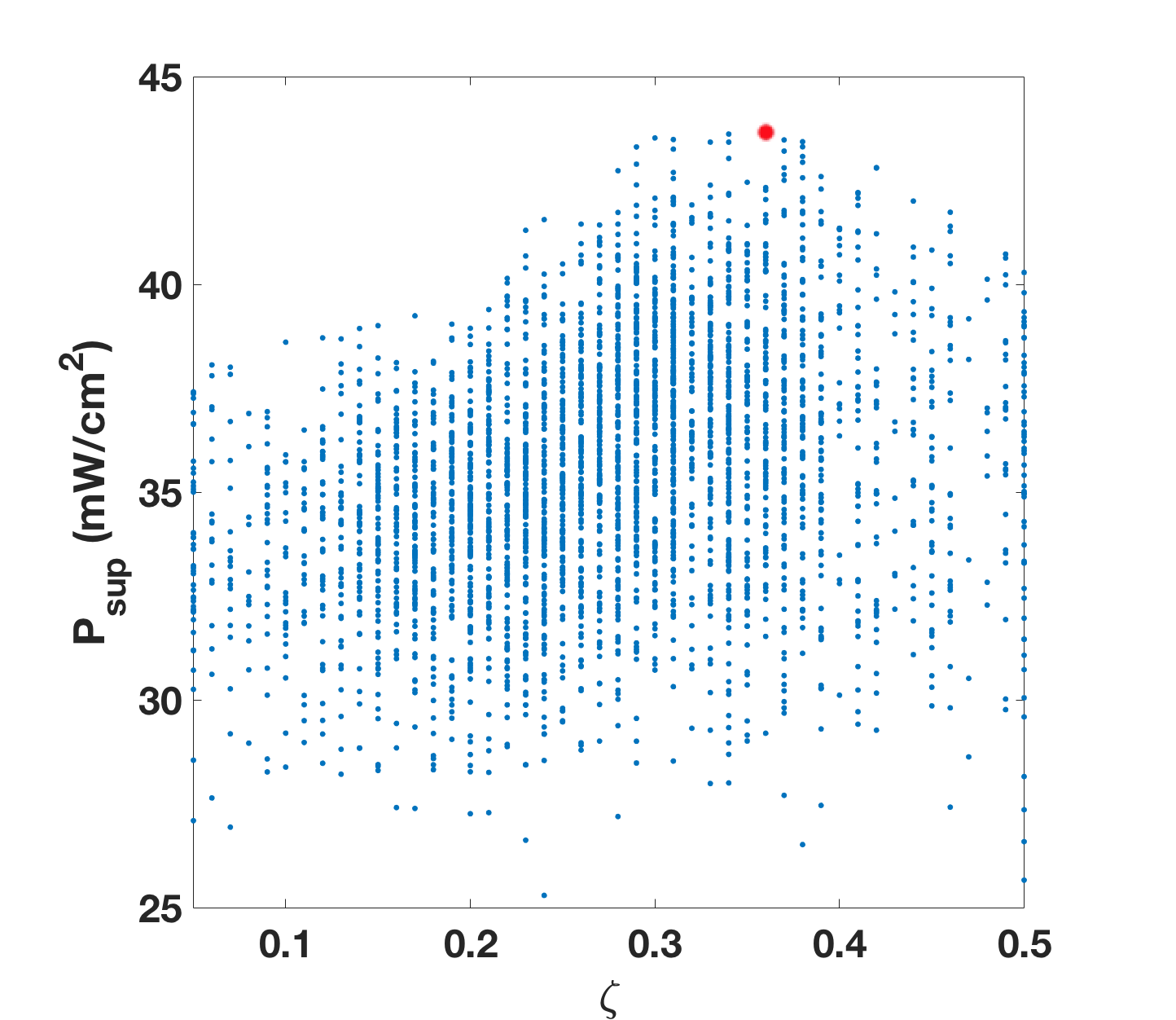} & \includegraphics[width=70mm]{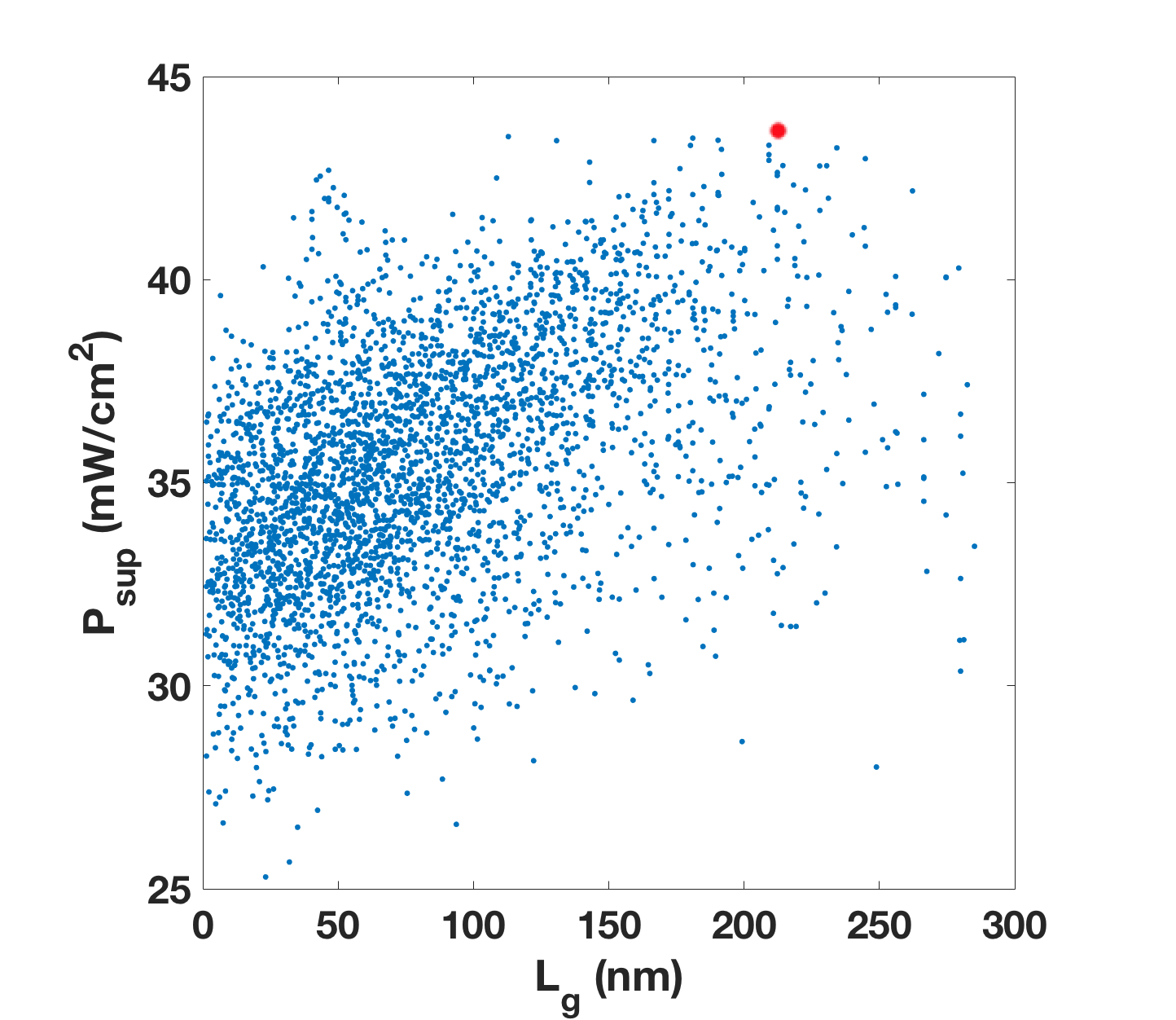} \\
(c) & (d) \\[6pt]
\end{tabular}
\end{figure}
\begin{figure}
\caption{\label{fig:DEA3} 
$P_{sup}$
in relation to (a) ${\sf E}_{1i}$, (b) ${\sf E}_{2i}$, (c) ${\sf E}_{3i}$, and $R_{g}$.
Each marker ($\mathbf{\cdot}$) represents a choice of parameters by DEA as the algorithm progresses. The red marker indicates the maximum value of $P_{sup}$ achieved. }
\begin{tabular}{cc}
  \includegraphics[width=70mm]{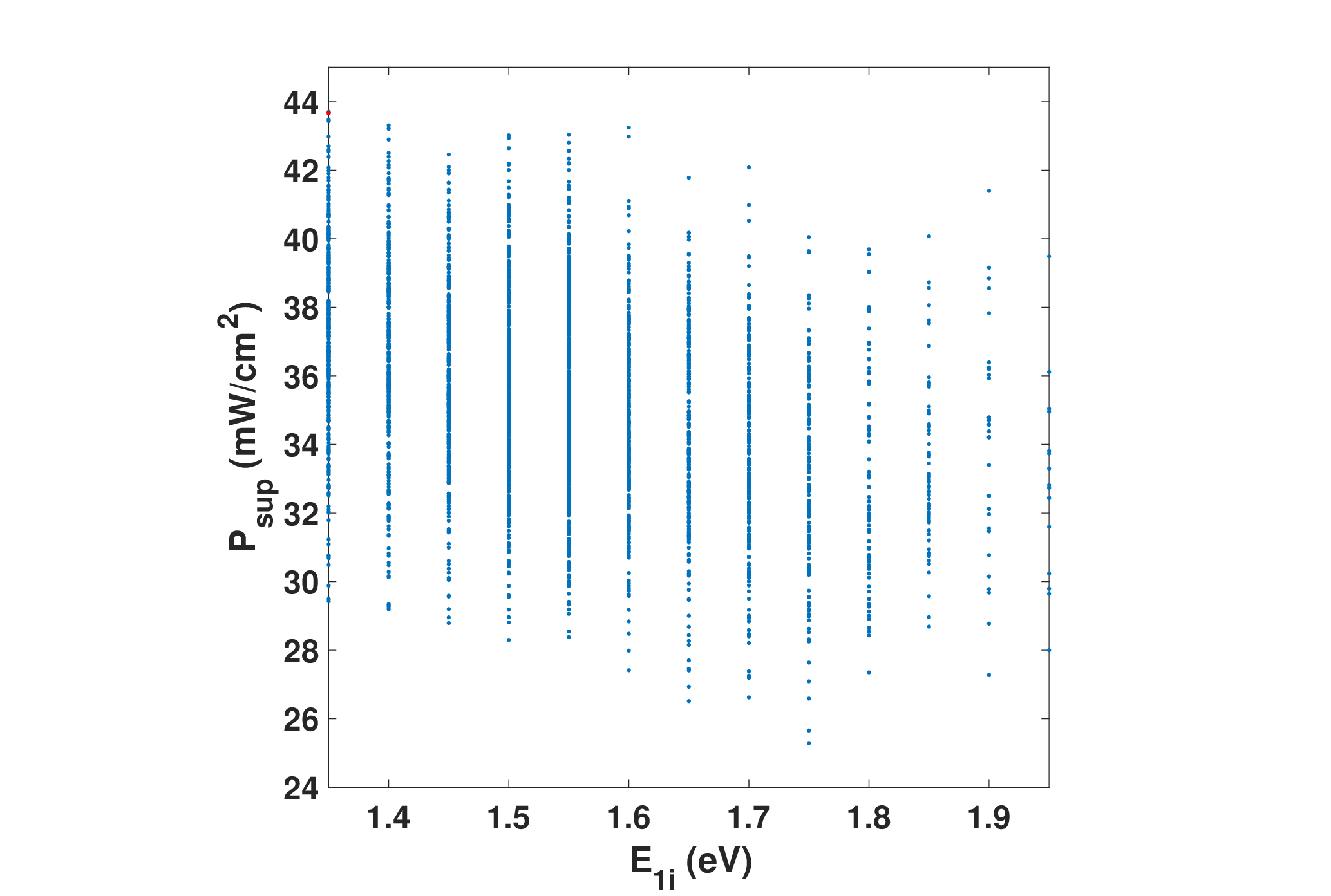} &   \includegraphics[width=70mm]{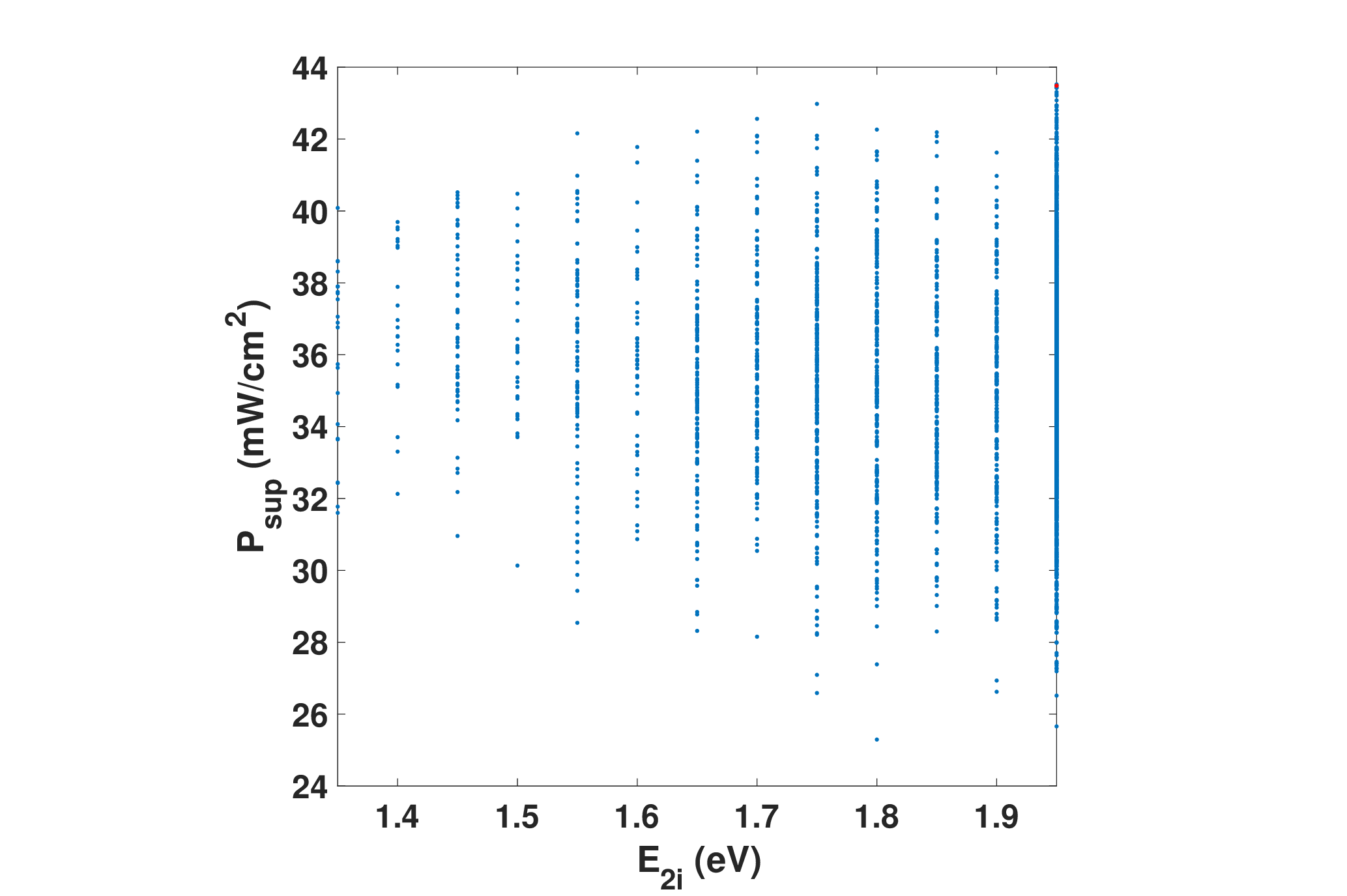} \\
(a)  & (b)  \\[6pt]
 \includegraphics[width=70mm]{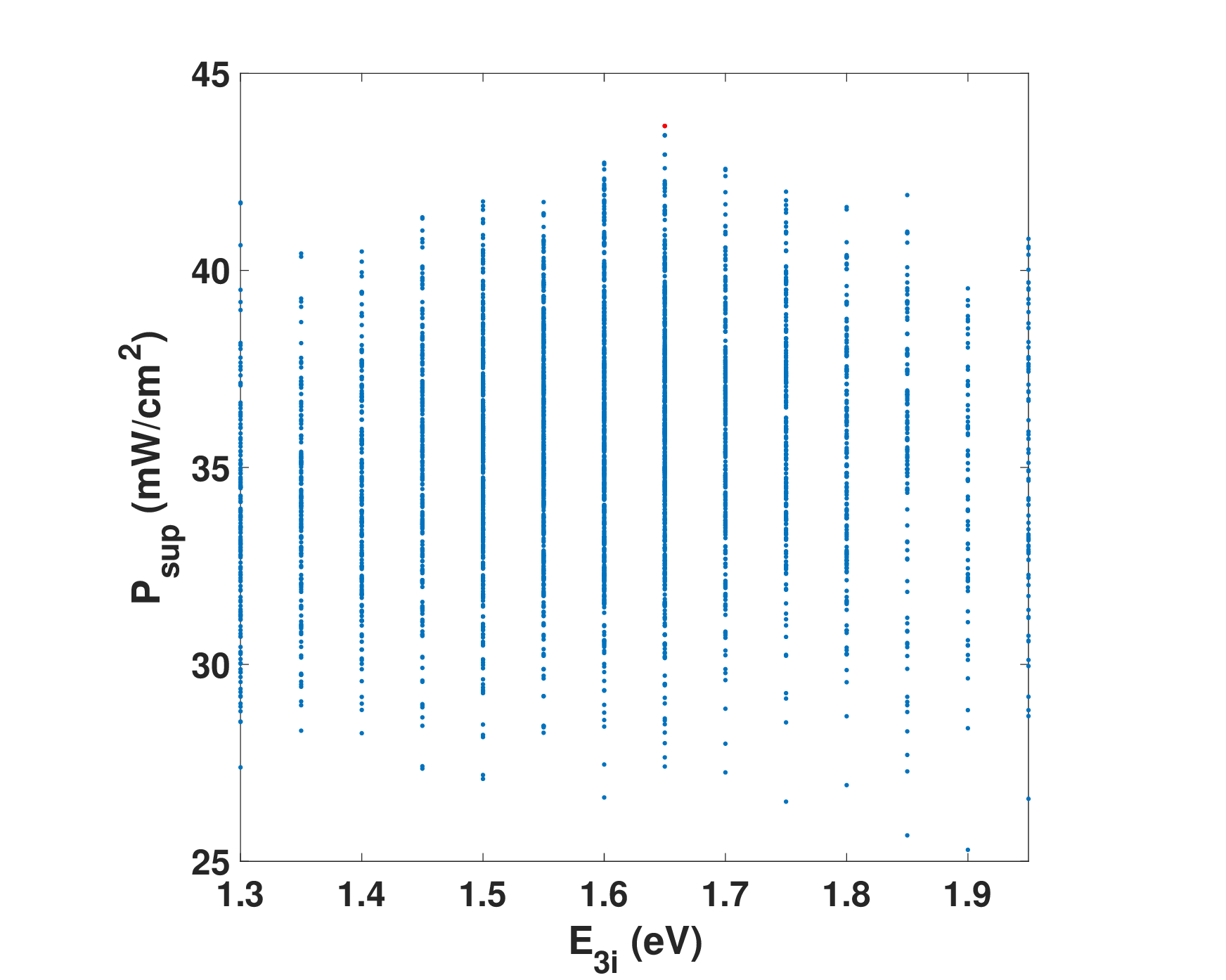} & \includegraphics[width=70mm]{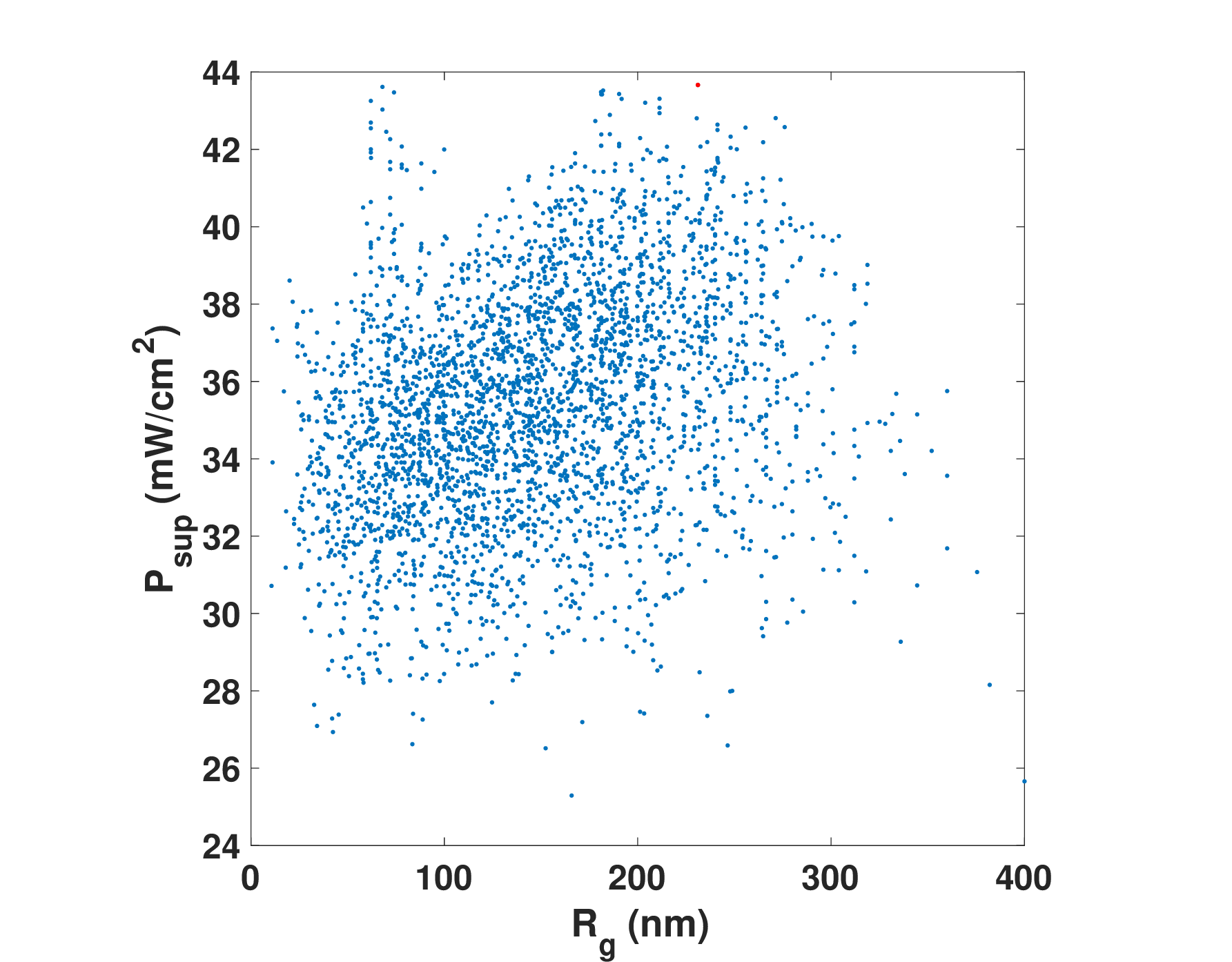} \\
(c) & (d) \\[6pt]
\end{tabular}
\end{figure}

Figures~\ref{fig:DEA2} and~\ref{fig:DEA3}  show the results of the DEA optimization for the   variable geometric and bandgap parameters of the 
solar cell---namely, the lattice parameter $L_h$, the antireflection-window thickness $d_{w}$, the duty cycle $\zeta={R_g}/L_h$, the corrugation height $L_{g}$,  and the bandgaps ${\sf E}_{1i}$, ${\sf E}_{2i}$, and ${\sf E}_{3i}$.  All seven of these parameters
influence the figure of merit $P_{sup}$ defined by Eq.~(\ref{eq:psup}).  Indeed, 
a steady increase in $P_{sup}$ is seen in
Fig.~\ref{fig:DEA2}(a) on the interval $200\leq L_h\leq 600$ nm;  a very sharp increase  in $P_{sup}$ is evident in
Fig.~\ref{fig:DEA2}(b) as $d_w$ increases from $10$ nm to $80$ nm;  $P_{sup}$ increases steeply  in
Fig.~\ref{fig:DEA2}(c) 
as $\zeta$ approaches $0.35$ with a drop off thereafter; and
$P_{sup}$ peaks in the neighborhood of $L_g=200$~nm in
Fig.~\ref{fig:DEA2}(d). Also shown is the hillock base radius $R_{g}$ in Fig.~\ref{fig:DEA3}(d). We note the contrast between the behavior seen in Fig.~\ref{fig:DEA1}(a) and Fig.~\ref{fig:DEA3}(c). In the latter optimization exercise, the effect of the other parameters amounts to a variation of $\pm 5$~mW~cm$^{-2}$. 

The  optimal values  found are as follows: ${\sf E}_{1i}=1.35$~eV, ${\sf E}_{2i}=1.95$~eV,
${\sf E}_{3i}=1.65$~eV,
 $L_h=642$~nm, $d_w=89$~nm, $\zeta=0.36$,
and  $L_{g}=231$ nm. 
These values yielded a maximum $P_{sup}=43.66$~mW~cm$^{-2}$.
These optimal parameters, found by maximizing $P_{sup}$, contrast sharply
with our findings in Sec.~\ref{Sec:4.1}, wherein maximization of $\JscOpt$
was dominated by the minimization of ${\sf E}_{3i}$ with the remaining $\Nbar-1$
parameters having very little effect.  

The optimization exercise yielded two distinct categories of unit cells with relatively high $P_{sup}$ values. The first comprises configurations for which $L_h$ lies in the interval $[200, 300]$~nm, and the second comprises configurations for which $L_h\in[550, 650]$~nm. For the first category, the optimal value of $L_{g}$ is approximately $40$~nm with a base radius $R_{g}$ of $50$~nm. The second type of configuration has much deeper corrugations, with $P_{sup}$ maximized when $L_{g}$ takes values near $200$~nm and $R_{g}$ around $230$~nm. This phenomenon is evidenced by two distinct peaks in Fig.~\ref{fig:DEA2}(a)--(d) and Fig.~\ref{fig:DEA3}(d). 

\subsection{Convergence of RCWA and DEA } 
\label{sec:converge}
\begin{figure}
\caption{\label{fig:DEA4} 
The best value of $P_{sup}$ versus number of DEA iterations when optimizing for maximum power density as in Sec.~\ref{sec:weight}.  }
  \includegraphics[width=70mm]{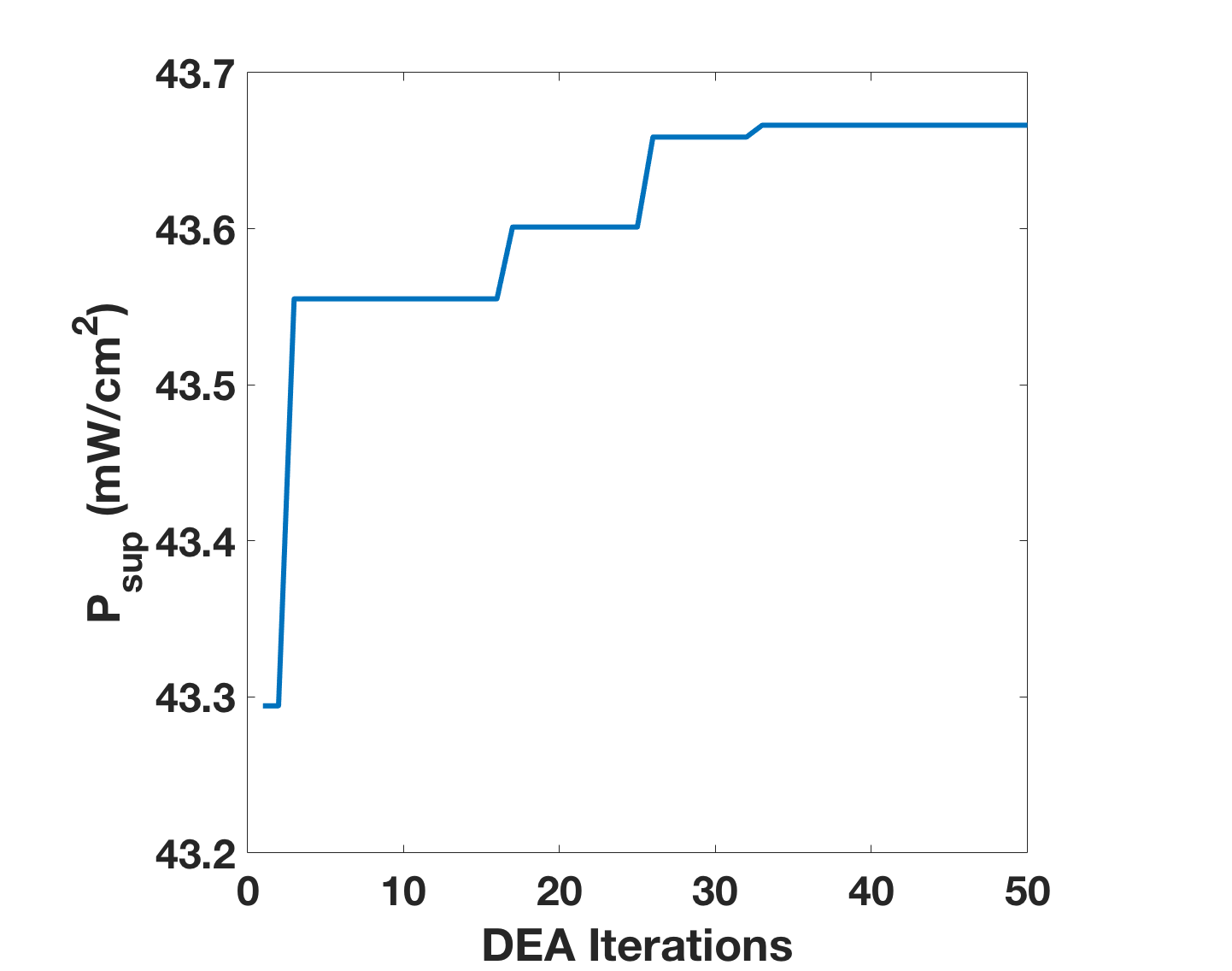} 
\end{figure}

To ensure convergence of the optical short-circuit density, a representative configuration for the unit cell was used to determine an appropriate choice of $N_t$ and $M_t$.  We let $N_t$ vary in the set $\{2,3,4,5,6 \}$, and defined $M_{t}= \lceil\sqrt{3} N_{t}\rceil $, where $\lceil \cdot \rceil$ is the ceiling function. After determining that  
 {$N_{ph}$} changed by $\leq 1 \%$ for two successive values of $N_{t}$ and $M_{t}$, the number of Fourier modes were fixed for all numerical  {results reported} in this paper, and taken to be $N_{t}=2$, $M_{t}=4$. Since Eq.~(\ref{eq:total}) has the charge-carrier generation rate integrated over the nine semiconductor layers, the susceptibility of $J_{SC}^{Opt}$ to the effect of 
 {Gibbs' phenomenon}\cite{gibbs} on the electric field in the region $z \in [0,L_{d}]$  is negligible. Hence, the electric field  {converges everywhere in the semiconductor layers}, even for relatively small values of $N_{t}$ and $M_{t}$. We performed 50 DEA iterations in the optimization exercise. Figure~\ref{fig:DEA4} shows the convergence of $P_{sup}$ with the number of DEA iterations. We see that the best value of $P_{sup}$ does not change after the first 32 DEA iterations. As with any
stochastic optimization method, it is always possible that further iteration would result in improvement of the computed maximum  {power density}.

\subsection{Comparison to bi-sinusoidal PCBR}
\label{sec:compare}
A third DEA-based optimization exercise was performed with the PCBR taken to be bi-sinusoidally corrugated \cite{faiz}. The relative permittivity $\eps_{g}(x,y,z,\lambdao)=\eps_{g}(x\pm L_x,y\pm L_y,z,\lambdao)$ in the grating region  is 
\begin{equation}
   \label{eq:perm1}
\eps_{g}(x,y,z,\lambdao)=
      \eps_{m}(\lambdao)-[\eps_{m}(\lambdao)-\eps_{w}(\lambdao)] \ \mathcal{U}(z-g_{1}(x)) \ \mathcal{U}(z-g_{2}(x)),\\
\end{equation}
with the corrugation-shape functions 

\begin{align}
g_{1}(x)&=L_{d}+L_{g}\big[1-\sin(2\pi\frac{x}{L_{x}})\big], \\
g_{2}(x)&=L_{d}+L_{g}\big[1-\sin(2\pi\frac{y}{L_{y}})\big]. 
\end{align}

The DEA parameters  were kept the same as in Sec.~\ref{sec:numerical}, and we set $L_x=L_y$ for the computation. The optimization exercise yielded an optimal configuration with $d_{w}=87$~nm, $L_{g}=110$~nm, $L_{x}=607$~nm, $R_{g}=212$~nm, ${\sf E}_{3i}=1.7$~eV, ${\sf E}_{2i}=1.55$~eV and ${\sf E}_{1i}=1.3$~eV. The maximal power density achieved was $P_{sup}=42.84$~mW~cm$^{-2}$, which is slightly lower than $43.66$~mW~cm$^{-2}$ obtained for the hexagonally corrugated
PCBR in Sec.~\ref{sec:weight}. However, the average $P_{sup}$ over all configurations visited by the DEA for
the bi-sinusoidally corrugated PCBR is  $38.17$~mW~cm$^{-2}$ but $35.06$~mW~cm$^{-2}$
for  the hexagonally corrugated
PCBR. There were many more configurations with relatively poor $P_{sup}$ values for the hexagonal case.

\section{Closing Remarks}
\label{sec:close}
As a model problem to demonstrate the practicability of the RCWA+DEA approach to design  {efficient 
 optically absorbing} 3D structures, optimal values of four geometric parameters and the bandgaps
of three $i$-layers were
found   for an  {idealized}, multi-terminal, thin-film tandem solar cell comprising
three $p$-$i$-$n$ junctions with a silver PCBR with  hillock-shaped corrugations arranged on a hexagonal lattice. 
  The figure of merit for the DEA was 
either (i) the optical short-circuit current density in Sec.~\ref{Sec:4.1} or (ii) 
the  power density in Sec.~\ref{sec:weight}.  Thus, two different optimization exercises using the RCWA+DEA
approach where performed.

As the optical short-circuit current density
takes into account only the optical constitutive properties of the solar cell,  {maximization of that
quantity resulted in a poor design. In particular,} we determined that  only the bandgap of the topmost $i$-layer (i.e., ${\sf E}_{3i}$) was significant to the optimization
of  the optical short-circuit current density, and the remaining six parameters played  minor roles. While  {photon} absorption in the topmost $p$-$i$-$n$ junction was maximized thereby, this configuration would have poor electrical performance.  {This is because}, when the thicknesses of the $i$-layers were included in the optimization, the DEA simply focused on the maximization of those  thicknesses.  Although  increasing those thicknesses can enhance light absorption, the electrical performance may be sacrificed, thereby reducing efficiency. The design of thin-film solar cells must balance optical and electrical performances \cite{optical,Anderson2016}.

In order to avoid configurations with potentially poor electrical performance, we next used the 
power density---thereby weighting the optical short-circuit current 
density---as a new figure of merit. The resulting configuration was optimal with respect to all seven design parameters. Another optimization exercise was then performed on a similar tandem solar cell, but with a bi-sinusoidally corrugated PCBR. In this case the configurationally averaged power density   tested by the DEA was about 10\% higher than with the hexagonally corrugated
PCBR. In the future, we plan to supplement the optical model by an electrical drift-diffusion model \cite{Anderson2016}  and then optimize the overall  {electrical} performance of the solar cell.

In closing, let us emphasize that the  triple-junction tandem solar cell was chosen as a model problem to show here the capabilities of the RCWA+DEA approach developed for 3D  {optically absorbing} structures. Our approach can be extended not only to other types of
photovoltaic solar cells \cite{Zhu,Kapadia}
but also to optical absorbers \cite{Liu,Butun} with 3D morphology.

\noindent{\bf Note.} This paper is substantially based on a paper titled, ``Optimization of charge-carrier generation in amorphous-silicon thin-film tandem solar cell backed by two-dimensional metallic surface-relief grating," presented
at the SPIE Optics and Photonics conference   Next Generation Technologies for Solar Energy Conversion VIII, held August 5--11,
2017 in San Diego, California, United States.\\

\noindent{\bf Acknowledgments.} The research of B.~J. Civiletti, T.~H. Anderson, and P.~B.  Monk is partially supported by  the US National Science Foundation (NSF) under grant number DMS-1619904.     The research of  F. Ahmed and A. Lakhtakia is partially supported by  the US NSF under grant number DMS-1619901. A. Lakhtakia thanks the Charles Godfrey Binder Endowment at the Pennsylvania State University for ongoing support of his research.

\end{document}